\documentclass{article}
\usepackage{graphicx}
\usepackage{color}
\usepackage{amsmath}
\usepackage{paralist}

\title{Breathers and kinks in a simulated crystal experiment}

\author{Q. Dou${}^1$, J. Cuevas${}^2$, J. C. Eilbeck${}^3$, and
  F. M. Russell${}^3$\\
  ${}^1$ School of Engineering and Physical Sciences\\Heriot-Watt
  University\\Riccarton, Edinburgh, EH14 4AS, UK\\
${}^2$ Grupo de F\'{\i}sica No Lineal.  Departamento de
  F\'{\i}sica Aplicada I.\\
 EU Polit\'ecnica.  Universidad de Sevilla\\
C/ Virgen de \'Africa, 7, 41011-Sevilla, Spain\\,${}^3$
Department of Mathematics and\\
Maxwell Institute for Mathematical Sciences\\
Heriot-Watt University\\
Riccarton, Edinburgh, EH14 4AS, UK}

\date{\today}

\begin{document}
\maketitle

\begin{abstract} 
  We develop a simple 1D model for the scattering of an incoming
  particle hitting the surface of mica crystal, the transmission of
  energy through the crystal by a localized mode, and the ejection of
  atom(s) at the incident or distant face.  This is the first attempt
  to model the experiment described in Russell and Eilbeck in 2007
  (EPL, {\bf 78}, 10004).  Although very basic, the model shows many
  interesting features, for example a complicated energy dependent
  transition between breather modes and a kink mode, and multiple
  ejections at both incoming and distant surfaces.  In addition, the
  effect of a heavier surface layer is modelled, which can lead to
  internal reflections of breathers or kinks at the crystal surface.

\end{abstract}

\section{Localized modes in Mica}
We describe elsewhere in these proceedings \cite{re09} the central
puzzle (what cause most of the macroscopic tracks in mica crystals) and our
interpretation (that these tracks are caused by quasi-particles called
quodons, which we postulate are longitudinal breathers).  In
particular, in this paper, we concentrate on the experiment described
in \cite{re07}.  In that paper, we describe a scattering experiment,
where alpha particles impinging on a mica crystal generate ejected
particles from the opposite face of the crystal.  Since the crystal in
question is of dimension of about $10^7$ lattice units, some
quasi-particle must have traversed this distance carrying sufficient
energy to eject an atom.  We call this quasi-particle a {\em quodon},
and postulate that it is a localized vibrational mode or breather.

In order to test this hypothesis, it is necessary to model the
experimental setup and obtain agreement between theory and
experiments.  It is believed that the K sheet in mica is responsible
for supporting such a breather, as the other planes are too rigid.
Our first step in this direction was a 2D simulation of breathers in
a hexagonal lattice representing the K sheet \cite{mer98}.  We found
evidence for mobile longitudinal optical breathers which traversed
about $10^4$ lattice units before breaking up.  Some recent studies by
Yi et al.\ \cite{ywsc09} have shown similar results in a somewhat
different lattice model.  If quodons are indeed breathers, we still
have a large factor of $10^3$ or more to account for in the comparison
between 2D models and the actual mica experiments and observations.
It may be that our model is insufficiently detailed or missing some
important feature, for example longer range interactions rather than
the nearest neighbour interactions assumed in the calculations.  It
may be that the full 3D case would introduce some new features which
would increase the breather lifetimes.  We plan to return to this
problem in the near future.

The calculations in \cite{mer98} were also limited with respect to
modelling the experimental situation in \cite{re07}, as no surface
effects were considered.  Our plan in this paper is to present a
modest 1D toy model which includes surface effects and hence is the
first step in modelling the recent experiments.  In all other respects
the model is a step backward, as it considers only 1D effects and
looks only at acoustic breathers (i.e. breathers with in-phase
oscillations).  Nevertheless it throws up some interesting new features
which may be of interest. In particular, as the energy of the incoming
particle increases, there is a transition in the internal localized
mode created, between a breather mode and a kink mode.  However this
transition is complicated by a third intermediate situation, where a
very slow or stationary breather is created, apparently pinned by the
lattice. Depending on the energy of the internal mode, multiple
ejections at both incoming and distant surfaces are possible.  In
addition, the effect of a heavier surface layer is modelled, which can
lead to internal reflections of breathers or kinks at the crystal
surface.

The paper is set out as follows.  We describe the model in the next
section, and Section 3 we briefly describe the results of various
simulations at increasing values of the energy of the incoming
particle.  In Section 4 we consider the interesting case where the
boundary particle(s) have a higher mass than those in the rest of the
crystal.  Finally some conclusions are given in Section 5.

\section{The model}

\begin{figure}
\includegraphics*[scale=0.5]{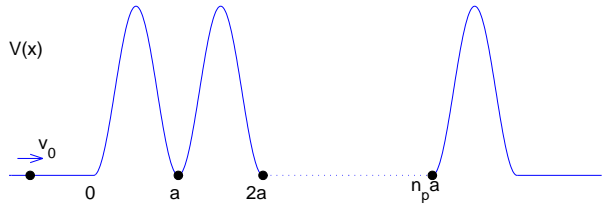}
\caption{Idealised model finite sized 1D mica crystal.  Note an
  incoming particle on the left about to strike the face of the crystal.}
\label{Fig1}
\end{figure}

Our idealised 1D model for the mica crystal experiment \cite{re07} is
shown in Fig \ref{Fig1}.
In the model, each atom in the crystal sits inside a harmonic lattice
potential, which is meant to model the effects of the adjacent sheets
in a higher dimensional crystal.  Outside the surface the potential is
set to be zero.  Note an incoming particle with velocity $v_0$ about
to strike the surface.  Particles are free to be ejected into the
vacuum on either side, from either the incident face or the distant
face, subject to the forces acting on them.  We assume all atoms
inside the crystal are at rest at a local minimum of a potential,
i.e.\ at absolute zero with no temperature effects.  Our Hamiltonian
is thus
\[
H = \sum_{i=0}^{n_p} \left[\frac12 \dot x_i^2+ V(x_i)+W(x_i-x_{i-1})\right],
\]
where the on-site potential $V$ is
\[
V(x)= \left\{ \begin{array}{ll} 0& \text{ if } x<0 \text{ or } x>L,\\
\frac{a^2}{4\pi^2}\left(1-\cos\left(\frac{2\pi x}{a} \right)\right) &
\text{ otherwise}. \end{array}
\right.
\]
In addition, the atoms are assumed to interact through an
inter-particle force given by a nearest neighbour Morse potential
$W(x_i-x_{i-1})$, where
\[
W(x)=\frac1{2b^2}\left[\exp(-b(x-a))-1\right]^2.
\]
Note the minimum of the Morse potential is assumed to coincide with
the lattice spacing $a$. However no correction is made at the surface
for the absence of a second particle to balance these forces, so our
model starts off unrelaxed with some small residual energy at the end
points.

As a further simplifying assumption, we assume the incoming particle
has the same (unit) mass and inter-particle force as the atoms resident
in the crystal.  We have modelled the surface potential in the simplest was
possible, by merely truncating the harmonic lattice potential to zero
at each end at the value corresponding to the minimum of the
sinusoidal potential.  Some alternative simulations with the external
constant potential corresponding to the maximum of the sinusoidal
potential have also been tried: in general the quantitative behaviour
in this case is very similar.

For simplicity we consider a small lattice of $n_p=64$ stationary
particles, plus one incoming particle.  In all our simulations we set
$a=1$ and $b=2$.  Except in the final section, all the masses in the
problem are normalized to unity.  The initial conditions are that
particles $1,\dots,n $ are at rest at $x_i=ia$, and particle 0 is
incoming at $-20a$ with velocity $v$.  The left hand boundary of the
crystal is at $x=0$ and the right hand boundary is at $x=L\equiv
(n_p+1)a$.  In order for $H$ to be defined for all $i=0,1,\dots,n_p$,
we introduce two fixed particles $-1$ and $n_p+1$ at $x=-1000a$ and
$x=1000a$ respectively.

An analysis of the linearised system shows that low amplitude coupled
oscillators with frequency below the phonon band must oscillate in
phase \cite{fl94,fg08}, so we do not expect optical breathers in this
system.  We refer the reader also to some previous studies of this
system, in which the interaction of breathers with both vacancies and
with interstitial defects \cite{ckaer03,casr06,cser09}.

\section{Dynamic simulations}
In this section we report a preliminary study with the model described
above, varying a single parameter, the velocity of the incoming
particle $v_0$.
\subsection{$v_0=0.4$}
\begin{figure}[h]
   \includegraphics*[scale=0.65]{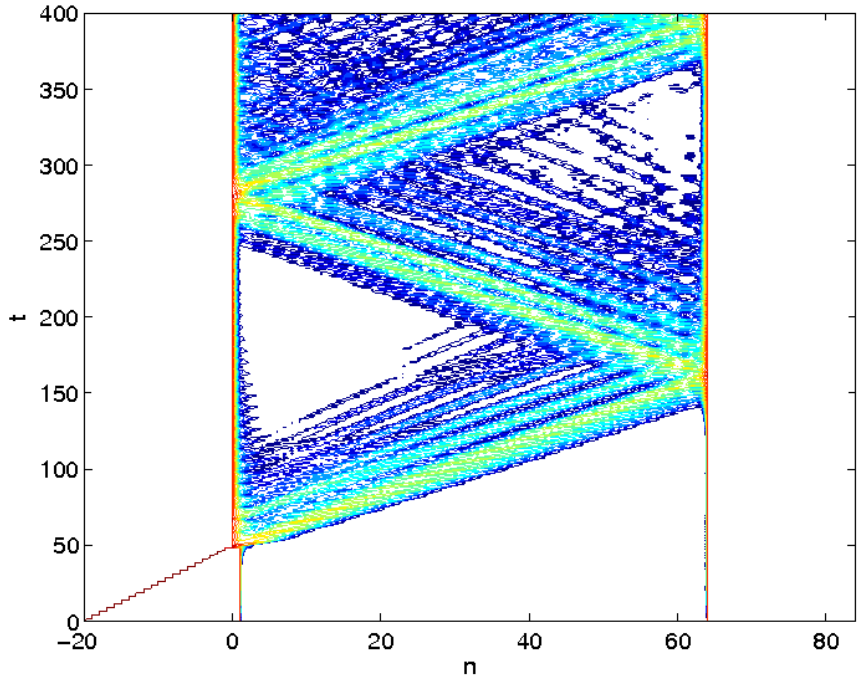} 
    \caption{Energy contours, $v_0=0.4$}
\label{FigV04}
\end{figure}

For low values of $v_0$, the incoming particle generates a low
amplitude fast acoustic breather which is reflected by the boundaries
but eventually begins to spread out and disperse.  Fig \ref{FigV04}
shows an example for $v_0=0.4$.
At these energies the displacement of the atoms in the lattice does
not show up clearly without some artificial scaling, so we have chosen
to display instead a contour plot of the local energy density.  Note
that in this figure and all subsequent energy plots, we are taking
contours of the logarithm of the energy.  This has the effect of
picking out small background effects, which might otherwise be swamped
by the main localized mode moving through the system.  We see a small
border region of high energy on each side due to the unrelaxed
boundary conditions.

\subsection{$v_0=0.65$}

As we increase $v_0$ further, the breather induced by the incoming
collision becomes slower and more energetic.  When it reaches the back
face of the crystal, the breather sometimes has sufficient energy and
a suitable phase to eject a particle.  The combination of energy and
phase is crucial, the final particle in the system will undergo a
strong induced oscillation, but this oscillation must be in phase with
the rest of the breather to accumulate enough energy for ejection.
Due to this effect we find a number of windows in the $v_0$ space
giving ejection, with other regions showing no effects except the
internal reflection of the breather at the surface.
  
\begin{figure}[h]
   \includegraphics*[scale=0.65]{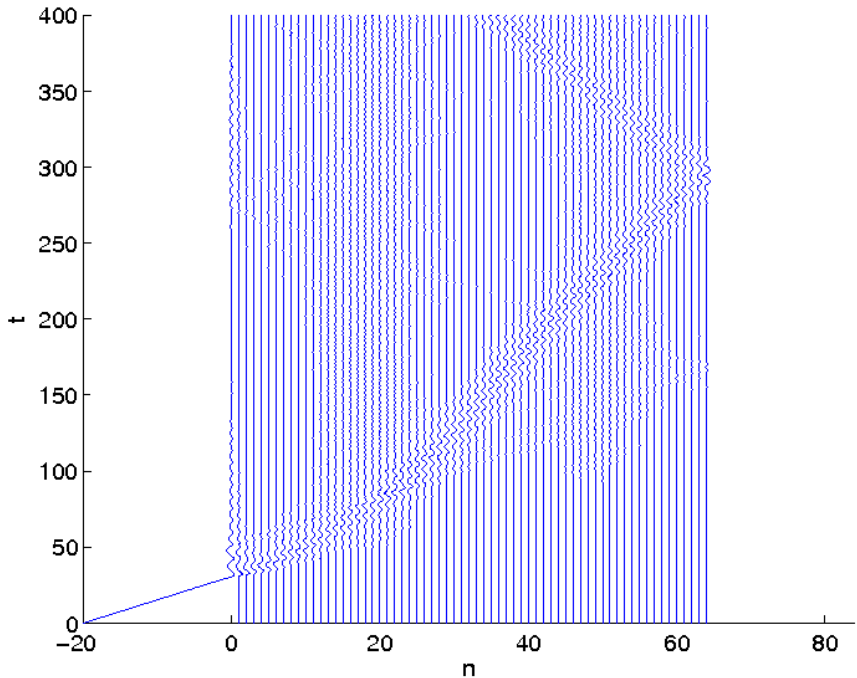} 
    \caption{Particle trajectories, $v_0=0.65$}
\label{Fig2}
\end{figure}

A typical example of the ``reflection only'' mode is shown in Fig
\ref{Fig2}.  Here we have chosen to display the particle trajectories
as a function of time for the case $v_0=0.65$.
The slanting line on the left shows the incoming particle, which
excites a mobile breather which then travels to the right.  The
incoming particle sticks to the surface, held there by the
inter-particle force between it and the first atom in the crystal.
Some of the energy is reflected in a smaller breather, and the rest
remains the back face, with the right-most atom making a small
excursion into the void on the right, but without enough energy to
escape completely.

\begin{figure}[h]
   \includegraphics*[scale=0.65]{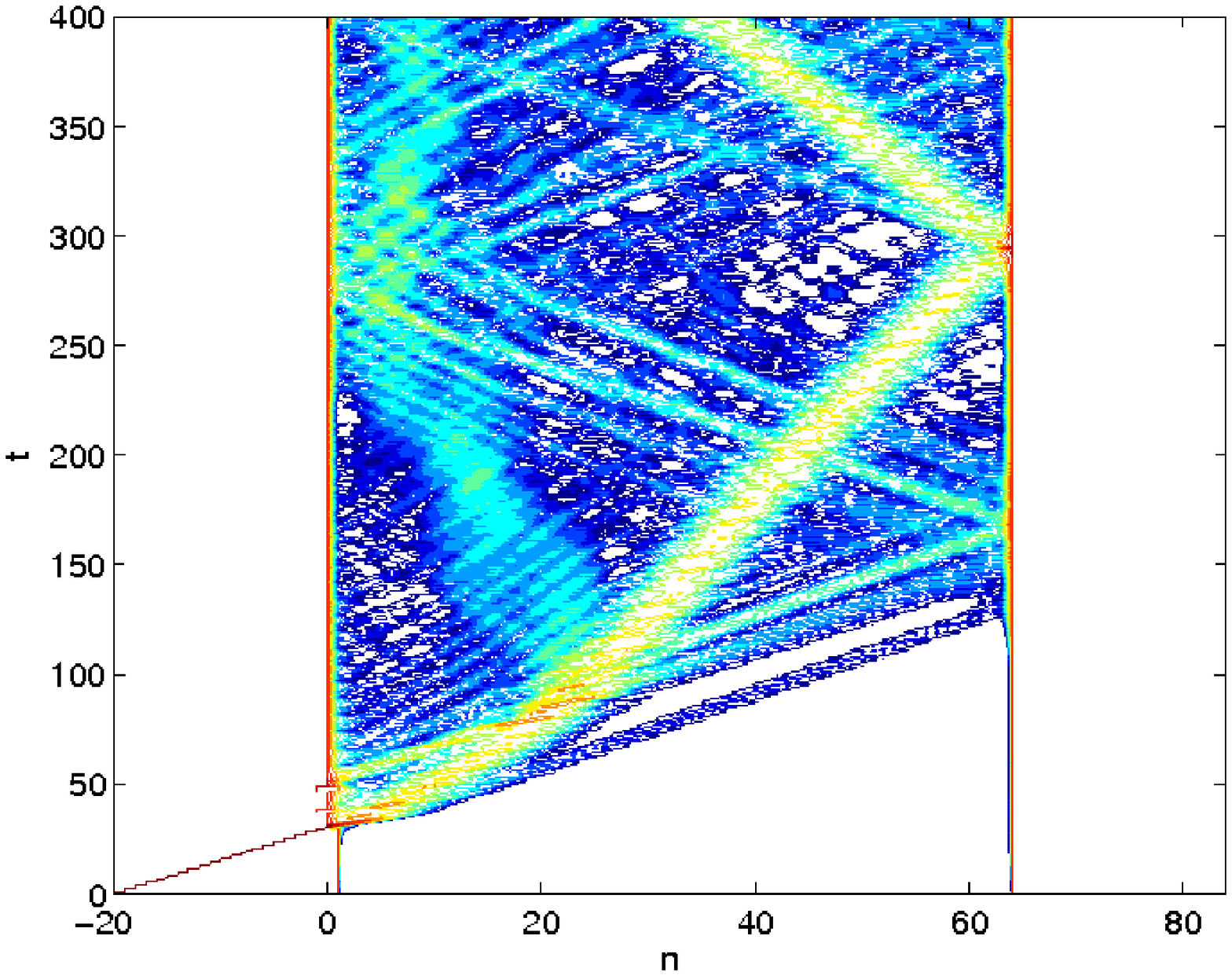}
   \caption{Energy contours, $v_0=0.65$}
\label{Fig3}
\end{figure}

The corresponding contour plot of the local energy density in this
case is show in Fig \ref{Fig3}, with the same parameters as Fig
\ref{Fig2}.
The main breather moving through the system can be clearly seen, as in
Fig \ref{Fig2}. However this plot shows some interesting fine detail
which is not clearly revealed in \ref{Fig2}.  A number of small
localized high speed modes are shown, together with a more extended
mode which breaks off from the main breather and travels slowly to the
left before colliding with the boundary and sticking loosely in this
position.  It would be interesting to study these modes further.

A further possibility, which we plan to investigate, is to fire a
sequence of atoms at the crystal.  As the energy builds up on the back
face, this will make it easier for subsequent breathers to eject
atoms from this ``hot-spot'' site.

\subsection{$v_0=0.66$.}
\begin{figure}[h]
   \includegraphics*[scale=0.7]{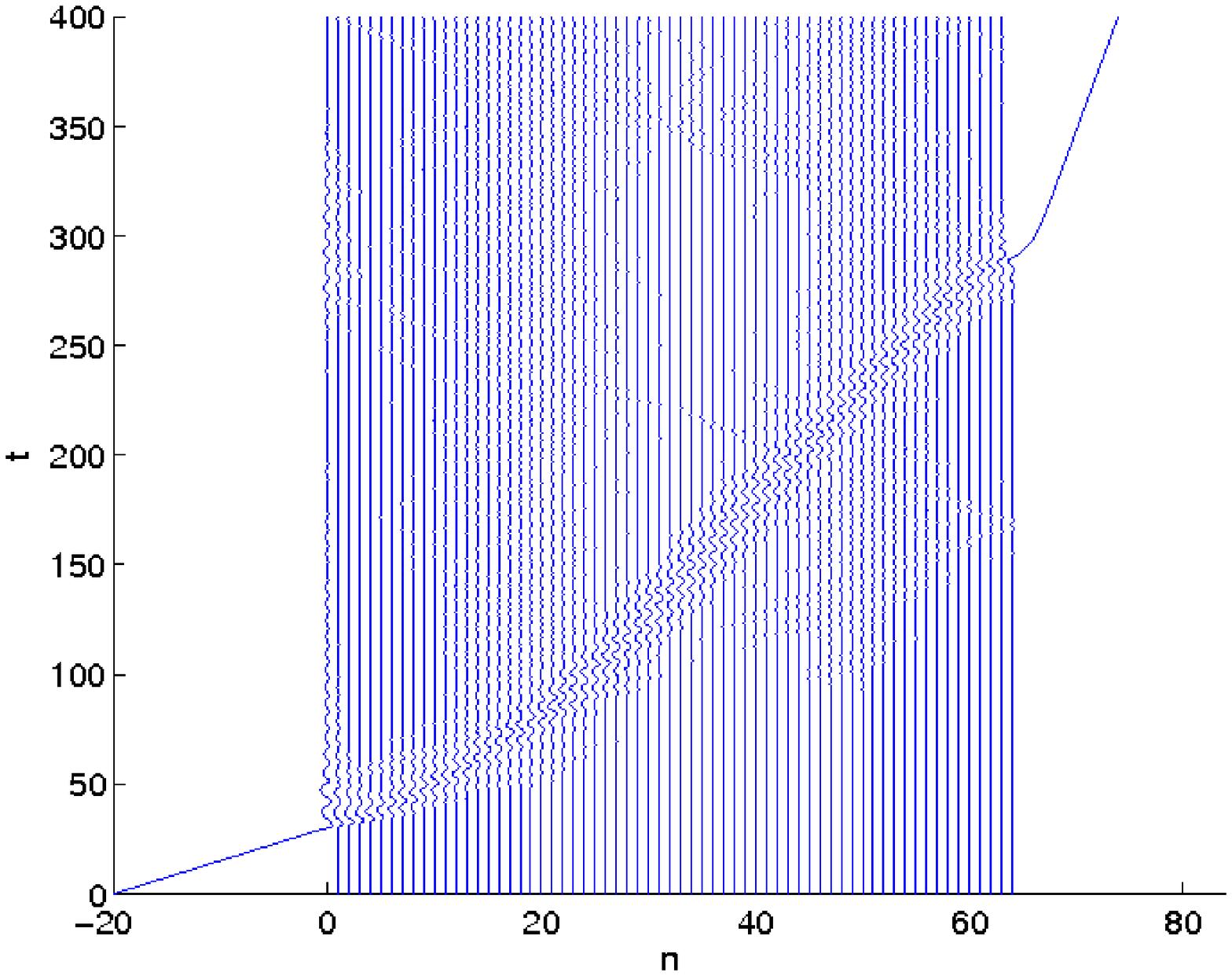} 
\caption{Particle trajectories, $v_0=0.66$}
\label{Fig4}
\end{figure}

If we increase $v_0$ slightly from from 0.65 to 0.66, we enter a
``ejection window'' and get the trajectory plot shown in Fig
\ref{Fig4}.
Although the creation and transmission of a breather proceeds as
before, when it reached the right-hand boundary, a particle is
ejected.  We emphasis again that this is not just because an energy
threshold for ejection has been reached, but because also the {\em
  phase} of the breather if favourable when the boundary is reached.
If for example the length of the crystal is increased slightly, the
phase of the breather may be different when it reached the boundary,
and no ejection is observed.  So it is not possible to predict whether
or not ejection takes place from knowledge of $v_0$ alone.  It would
be interesting to examine a detailed range of $v_0$ in this region to
chart the different windows -- we have observed at least three, but
there may be many more.

\subsection{$v_0=0.8$}
\begin{figure}[h]
   \includegraphics*[scale=0.65]{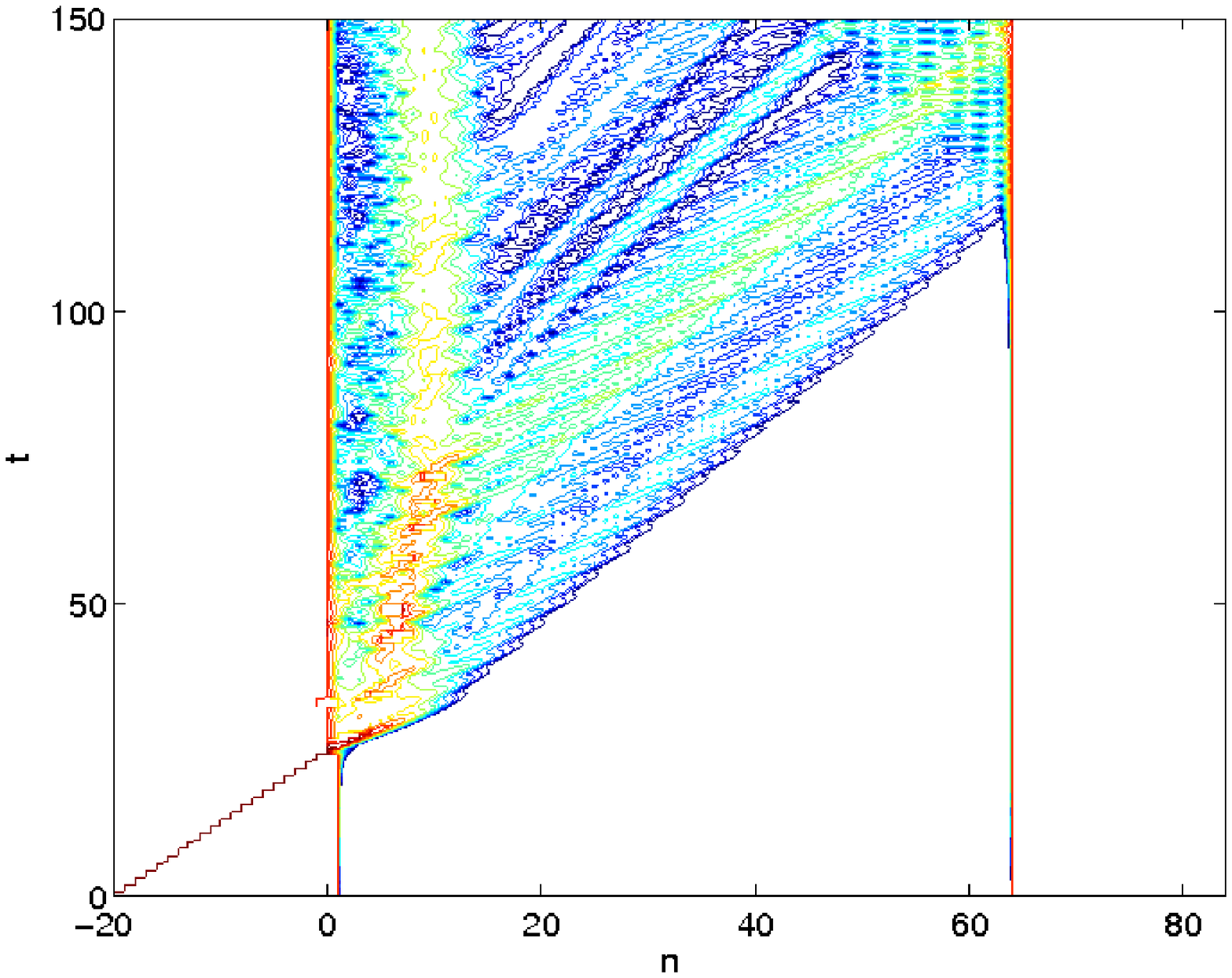} 
\caption{Energy density plot, $v_0=0.8$}
\label{Fig5}
\end{figure}

Naively one might think that if we continue to increase the value of
$v_0$, the energy of the induced breather is increased accordingly,
and ejection is more likely at the far edge of the crystal.  However
it has long been observed that at higher energies, the discreteness of
the lattice plays an increasingly important role in breather
propagation, and eventually we get a ``pinning'' effect.  In the
current model this is indeed observed.  The effect is most easily seen
with an energy contour plot.  If we increase $v_0$ to 0.8, we get the
result shown in Fig \ref{Fig5}.
When the incident particle hits the crystal, a stationary localized
mode is formed which after a delay throws off some small high-speed
breathers. The bulk of the energy stays trapped close to the l.h.\
boundary, although some small oscillations in its position are seen.
However we have again observed sensitive behaviour dependent on
initial conditions - for a small change in $v_0$ the velocity and
trajectory of the induced breather can be quite different.

\subsection{$v_0=0.88-0.89$}
Ad $v_0$ increases further, we enter a complicated region with results
even more sensitive to the specific value of $v_0$ used.  We get the
first examples of kinks occurring simultaneously with mobile breathers,
also stationary surface breathers at the incident surface which can
lead to ejections from this face.  A nice example showing all three
effects together is shown below in Fig \ref{Fig5a} for $v_0=0.887$,
which should be studied in conjunction with the corresponding energy
density plot in Fig \ref{Fig5b}. 
\begin{figure}[h]
   \includegraphics*[scale=0.65]{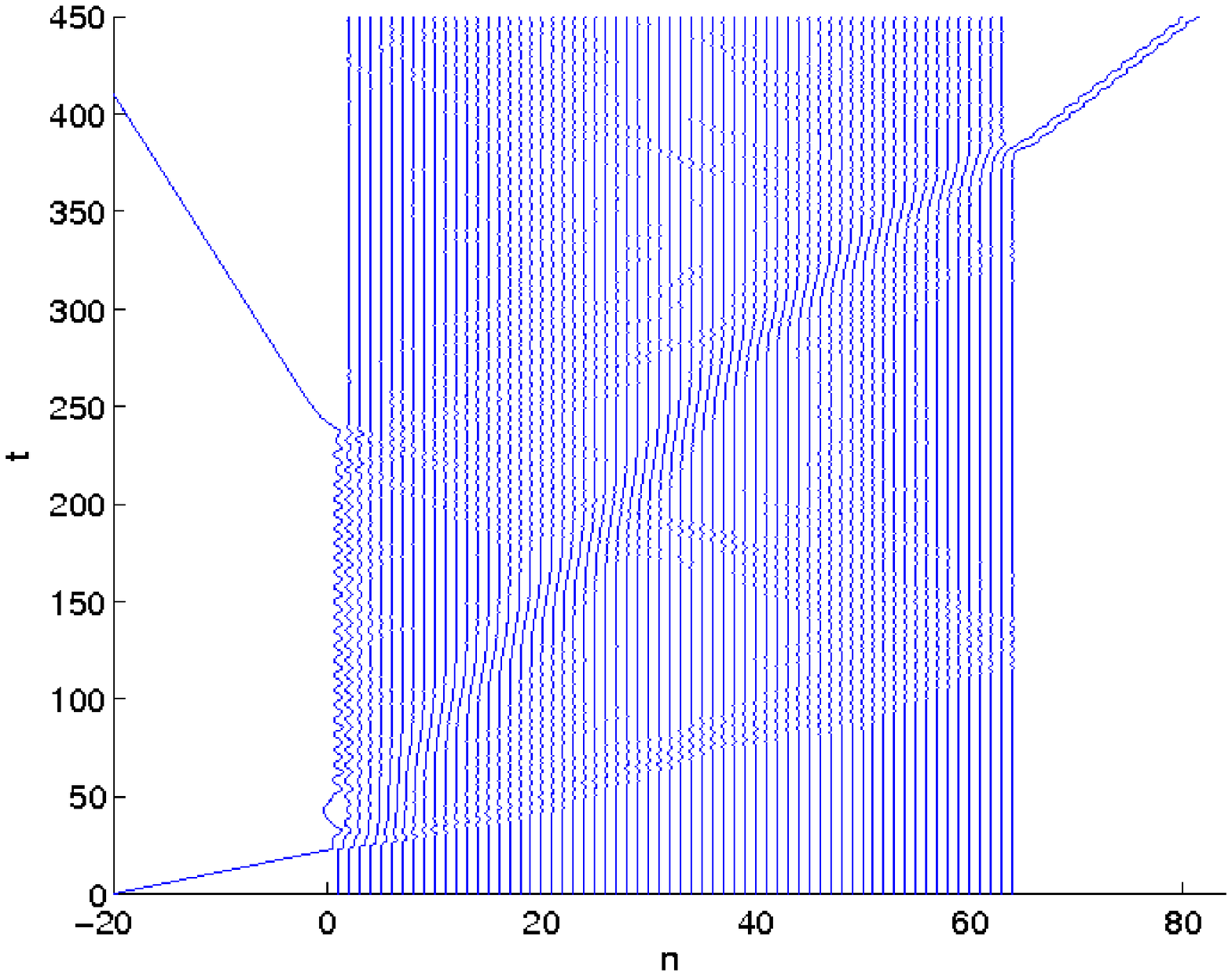} 
\caption{Particle trajectories, $v_0=0.887$}
   \label{Fig5a}
\end{figure} 
\begin{figure}[h]
   \includegraphics*[scale=0.65]{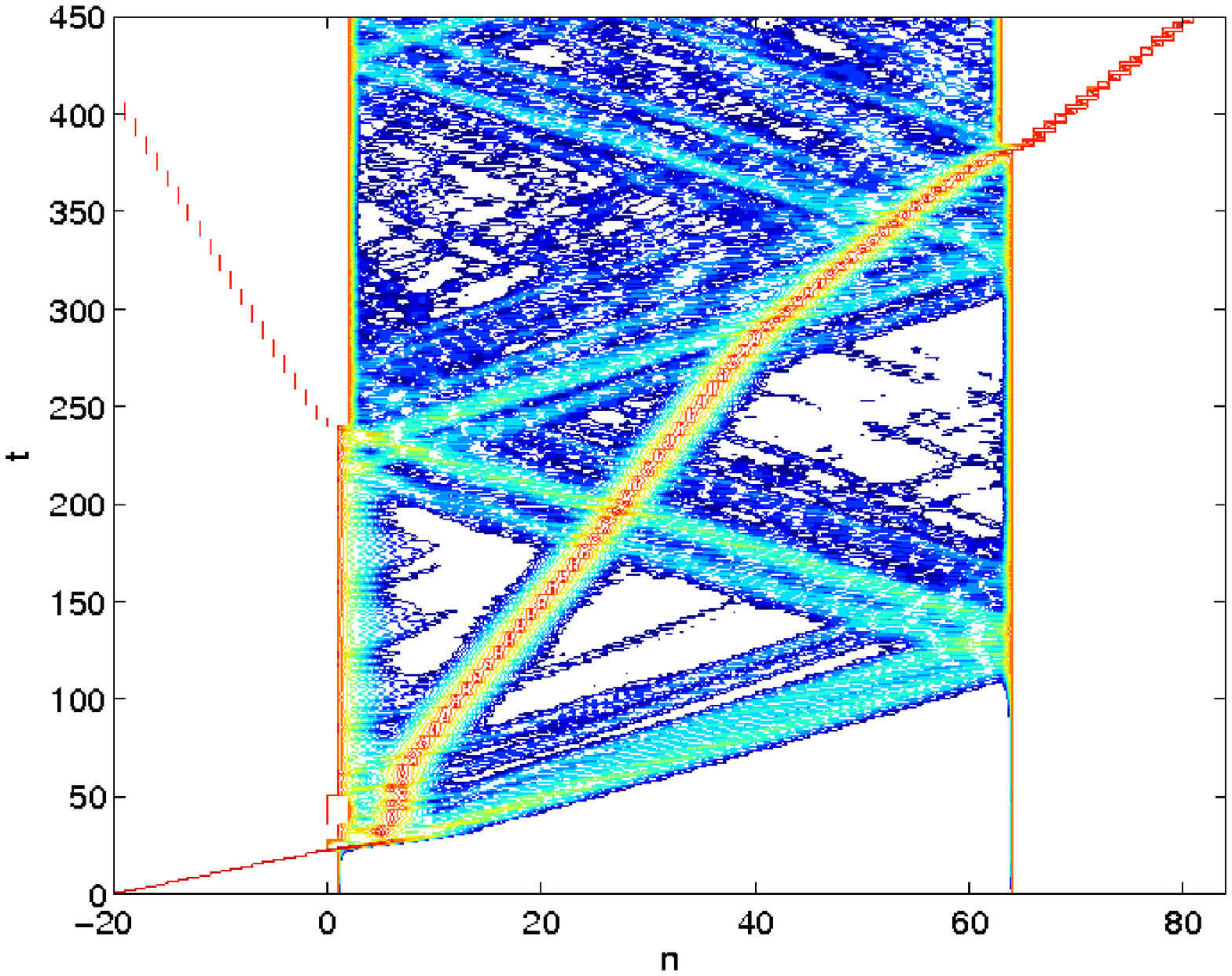} 
\caption{Energy density plot, $v_0=0.887$}
   \label{Fig5b}
\end{figure}
In this example the incoming particle interacts in a complicated way
with the atoms in the incident surface.  A boundary excitation is set
up which persists for a long time.  Meanwhile a small fast breather is
generated which travels to the far edge and is reflected.  When it
returns to the incident surface, it has sufficient energy to nudge the
surface mode into ejecting an atom to the left.  Note that if our
model included internal vacancies or interstitials, we would expected
that in at least some cases the mobile breather would be reflected by
this impurity, and a similar effect might be induced.  In sputtering
experiments a delayed ejection of this sort has often been seen, in
specific crystal axis directions, but the suggested mechanism is very
different.

In addition to all this a travelling topological kink is formed which
moves more slowly from left to right, and ejects two particles at the
right-hand face.  By kink we mean a local mode in which the atom in a
particular potential well suffers a permanent translation to the right
or left as the mode travels through.

Other choices of $v_0$ in the range $v_0=0.88-0.89$ gives a variety
of combinations of phenomena, including immediate ejection of atoms
from the incident face, formation of vacancies just inside the
incident face, single or double ejections from the far face, etc.
 
\subsection{$v_0=1.05$}
\begin{figure}[h]
   \includegraphics*[scale=0.65]{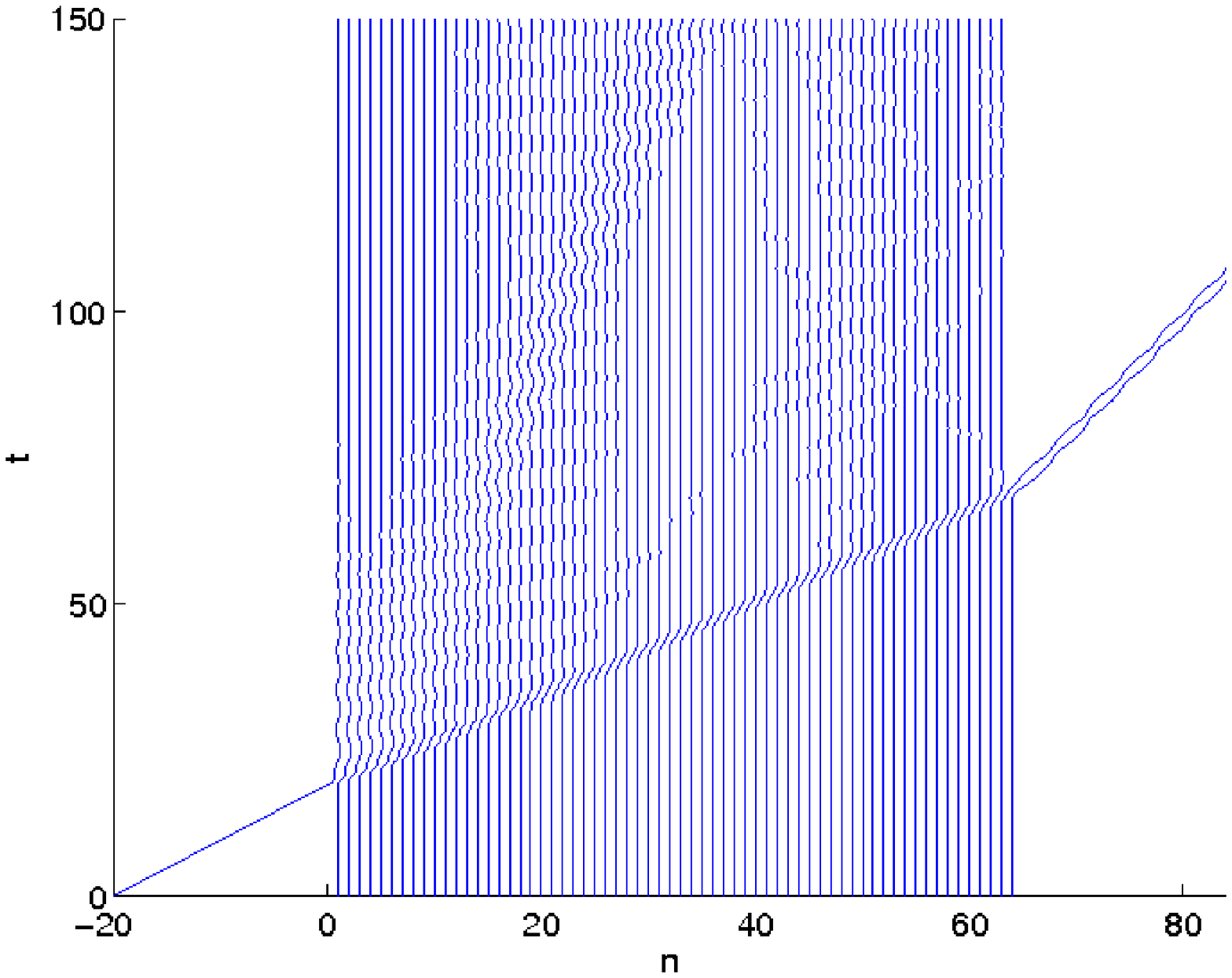} 
\caption{Particle trajectories, $v_0=1.05$}
   \label{Fig6}
\end{figure}

If we further increase $v_0$ to 1.05, the trapping phenomena switches
to pure kink behaviour, with no significant breather activity.  This
is illustrated in Fig \ref{Fig6}.
Now the incoming particle induces a more energetic topological kink.
There is little evidence of a faster breather, as in Figs
\ref{Fig5a}, \ref{Fig5b}.  When the kink reaches the right-hand boundary, as
in the previous case, it has sufficient energy to eject two particles,
which stay close enough together to be bound by the inter-particle
potential.  In addition some energy is deposited in a breather close
to the left hand boundary, where it moves slowly to the right in
contrast to the one shown in Fig \ref{Fig5}.

\subsection{$v_0=1.65$}
\begin{figure}[h]
   \includegraphics*[scale=0.65]{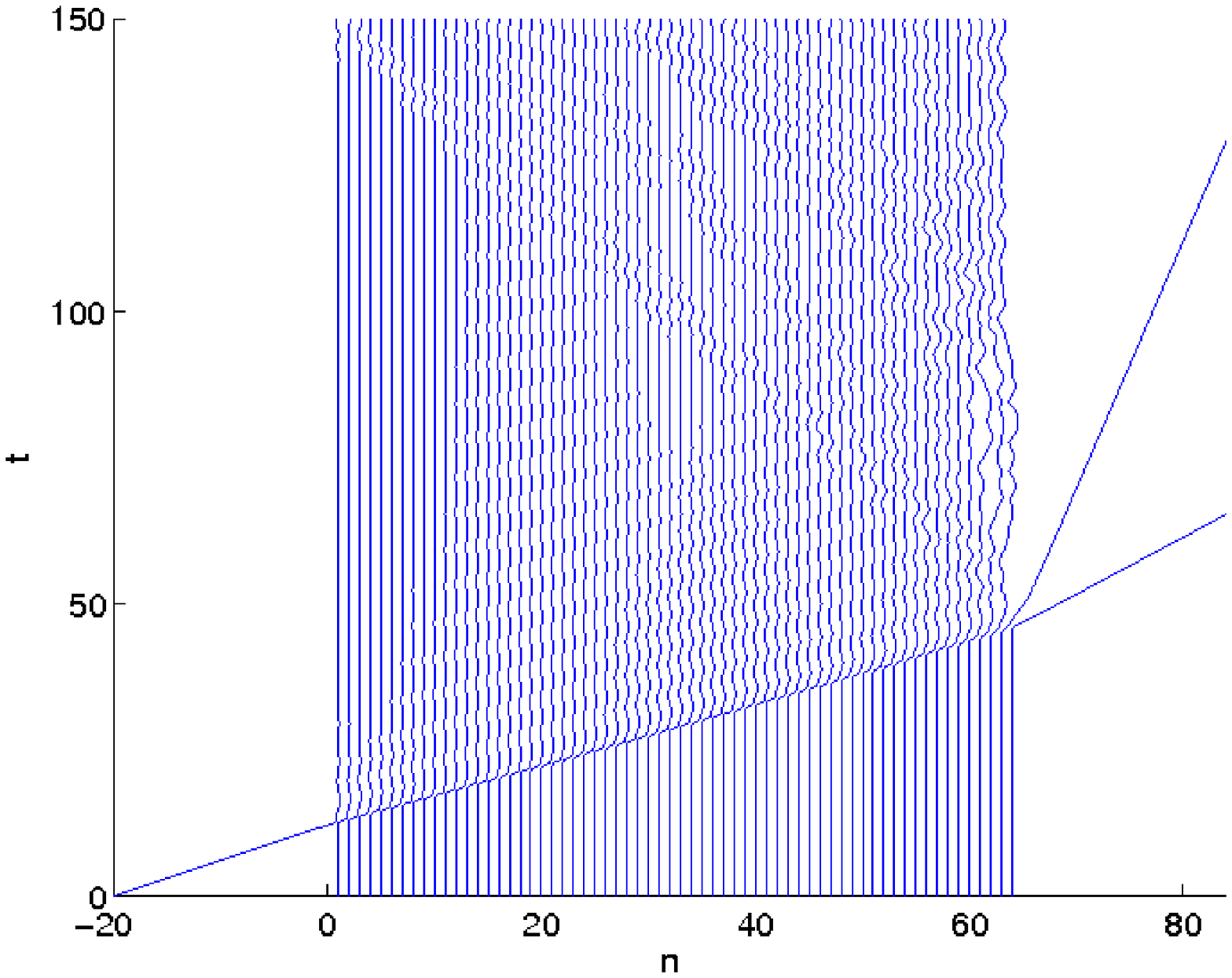}   
\caption{Particle trajectories, $v_0=1.65$}
   \label{Fig7}
\end{figure}

A further increase in $v_0$ does not apparently produce any new
internal localized mode, it seems always the case that kink is formed,
although the higher the energy, the narrower the kink appears.  Fig
\ref{Fig7} shows a calculation with $v_0= 1.65$.  The main qualitative
difference with the previous diagram is that in this case the two
ejected particles are no longer bound together but are thrown off with
two different velocities.  In addition, a lot of energy is left at the
boundary after ejection, some of which is reflected as one or more
breathers.

\section{Higher mass boundary layers}
\begin{figure}[h]
   \includegraphics*[scale=0.65]{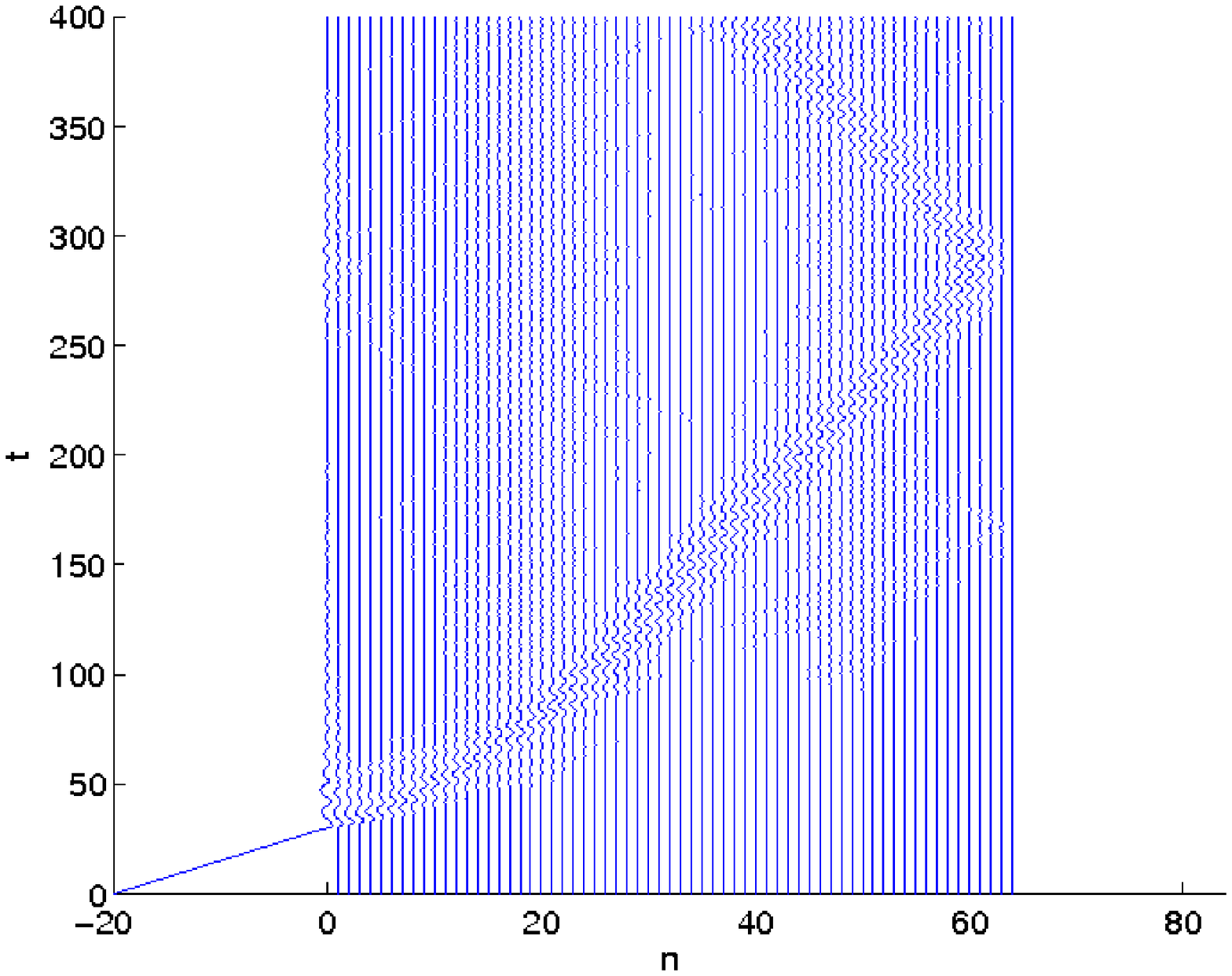} 
\caption{Particle trajectories, $v_0=0.66$, r.h.\ mass 10 units}
 \label{Fig8}
\end{figure}

In some applications \cite{re09}, we may wish to reduce the
possibility of ejection and increase the amount of energy that
undergoes internal reflection at a boundary.  This could enable us to
have a much longer breather lifetime in a small crystal.  A good
approximation to fixed rigid boundary conditions is to increase the
mass of one or more of the boundary atoms in the crystal.  This could
be effected in practice by depositing a layer of heavy atoms on the
surface by sputtering.  The effect of having a single atom on the
right hand boundary, with 10 times the mass of the other atoms in the
crystal, is shown in Fig \ref{Fig8} for the case $v_0=0.66$.  
This should be compared with Fig \ref{Fig4} for the single mass case.
It is clear that ejection is suppressed and more energy is reflected.
In view of the sensitivity of the ejection process to phase, etc., we
should report that this feature seems robust in several test runs with
similar parameters.

This situation would also model approximately the case of a heavy
impurity at an internal site in the crystal, if impact energies were
low.

\begin{figure}[h]
  \includegraphics*[scale=0.65]{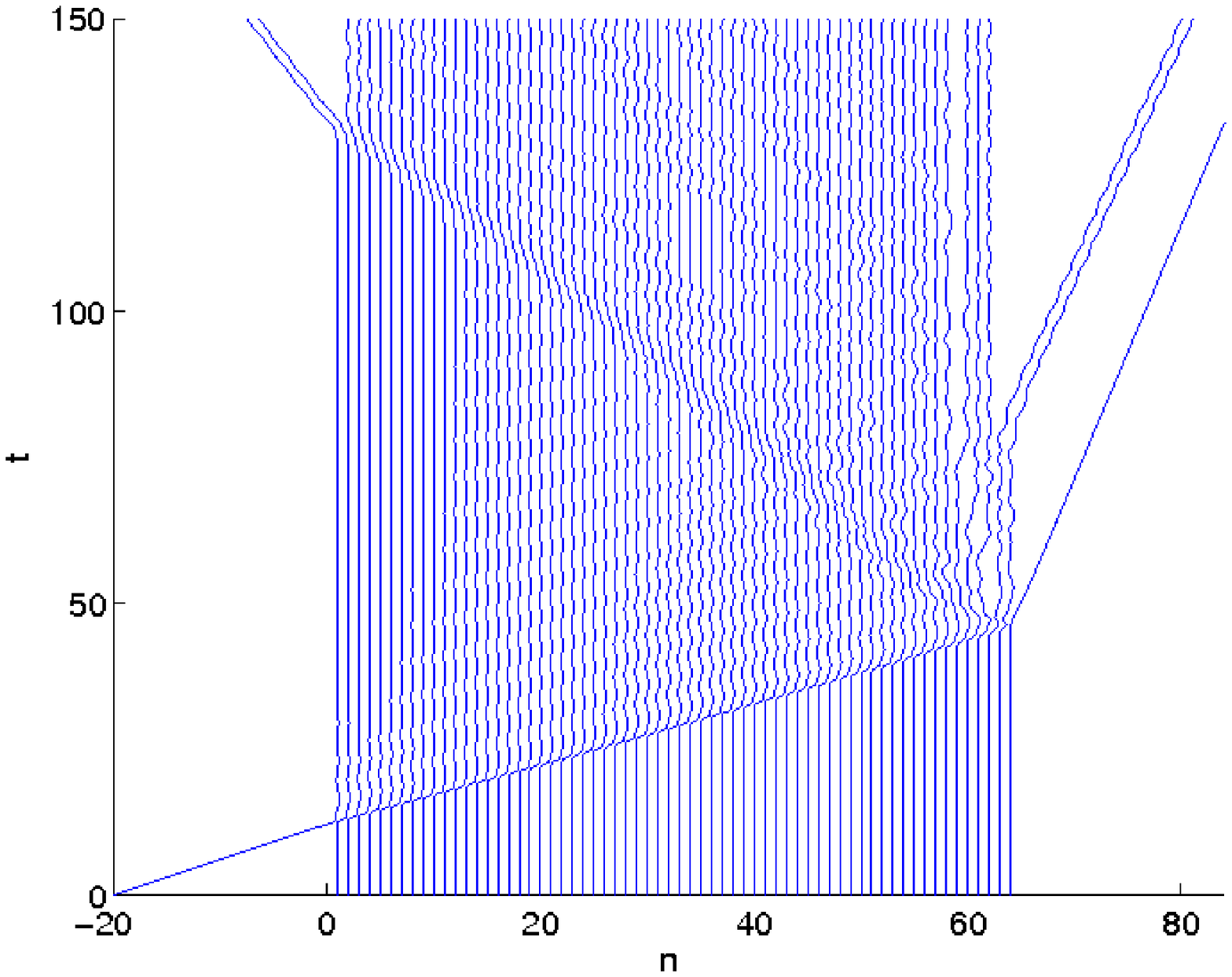} 
\caption{Particle trajectories, $v_0=1.65$, r.h.\ mass 10 units}
 \label{Fig9}
\end{figure}

However, once we go to high energies, we can get very different and more
complicated features.  Fig \ref{Fig9} shows this for the case
$v_0=1.65$.
This should be compared with Fig \ref{Fig7} for the single mass case,
and also Figs \ref{Fig5a}, \ref{Fig5b}.  Now there is an initial
ejection at the r.h.\ face when the initial kink hits, but some of the
energy is reflected in a slower kink which in turn ejects two
particles from the r.h.\ face when it arrives.  Meanwhile much energy
is left in a complicated stationary mode at the r.h.\ face.  Some of
this results in a reflection of a less energetic kink which in turn
ejects two particles from the l.h.\ face when it arrives.  The
remaining energy pinned at the r.h.\ face eventually ejects two more
particles after a delay, leaving an embedded vacancy (stationary
anti-kink).  In contrast to Figs \ref{Fig5a}, \ref{Fig5b}, these high
energy effects are qualitatively more robust to small changes of
$v_0$.  Note that the ejection at the l.h.\ face can be considered as
a delayed sputtering, and could equally well be caused by reflection
from an embedded impurity in the crystal.  We have already remarked on
the delayed sputtering phenomena in our discussion of Figs
\ref{Fig5a}, \ref{Fig5b}.

\section{Conclusions}
Our preliminary study shows that we can induce acoustic (in phase)
breathers by scattering and these can eject atoms in some cases.  As
the energy of the incoming particle increases, there seems to be a
transition from mobile breathers to trapped breathers, then a further
transition from trapped breathers to mobile kink as energies increase.
These transitions are complicated by the fact that particle ejection
and other phenomena depend on the phase of the induced breathers as
well as their amplitudes.  In addition there is very sensitive
response to initial conditions due to the various nonlinear resonances
in the system.  We further show that heavy atom(s) at boundaries can
simulate fixed boundary conditions and increase reflection of energy.
Our work has thrown up a number of interesting effects deserving more
systematic study in the future.  For example we need to map out in
detail the different phenomenon ranges, and the properties of
generated kinks and breathers after the initial impact and after
collisions with distant boundaries.  The development of the initial
motions on impact also requires further investigation.  The effect of
a sequence of incident particles rather than just one is also worth
studying.

\section*{Acknowledgement}
The support given by Turbon International Limited is
acknowledged.


\begin{thebibliography}{1}

\bibitem{re09}
F.~M. Russell and J.~C. Eilbeck.
\newblock Persistent mobile lattice excitations in a crystalline insulator.
\newblock To be published in the proceedings of the conference
  ``Localized Excitations in Nonlinear Complex Systems (LENCOS)'',
  Sevilla (Spain) July 14-17, 2009.

\bibitem{re07}
F.~M. Russell and J.~C. Eilbeck.
\newblock Evidence for moving breathers in a layered crystal insulator at 300k.
\newblock {\em Europhysics Letters}, 78:10004, 2007.

\bibitem{mer98}
J.~L. Mar\'{\i}n, J.~C. Eilbeck, and F.~M. Russell.
\newblock Localized moving breathers in a 2-{D} hexagonal lattice.
\newblock {\em Phys. Letts. A}, 248:225--229, 1998.

\bibitem{ywsc09}
X.~Yi, J.~A.~D. Wattis, H.~Susanto, and L.~J. Cummings.
\newblock Discrete breathers in a two-dimensional spring-mass lattice.
\newblock {\em J. Phys. A}, 42:355207, 2009.

\bibitem{fl94}
S.~Flach.
\newblock Conditions on the existence of localized excitations in nonlinear
  discrete systems.
\newblock {\em Phys. Rev. E}, 50:3134, 1994.

\bibitem{fg08}
S.~Flach and A.~Gorbach.
\newblock Discrete breathers —- advances in theory and applications.
\newblock {\em Phys. Rep.}, 267:1, 2008.

\bibitem{ckaer03}
J.~Cuevas, C.~Katerji, J.~F.~R. Archilla, J.~C. Eilbeck, and F.~M. Russell.
\newblock Influence of moving breathers on vacancies migration.
\newblock {\em Phys. Lett.}, 315:364--371, 2003.

\bibitem{casr06}
J.~Cuevas, J.F.R. Archilla, B.~Sánchez-Rey, and F.R. Romero.
\newblock Interaction of moving discrete breathers with vacancies.
\newblock {\em Physica D}, 216:115, 2006.

\bibitem{cser09} J.~Cuevas, B.~Sanchez-Rey, J.~C. Eilbeck, and F.~M.
  Russell.  
  \newblock Interaction of moving discrete breathers with simulated
  interstitial defects.
  \newblock To be published in the proceedings of the conference
  ``Localized Excitations in Nonlinear Complex Systems (LENCOS)'',
  Sevilla (Spain) July 14-17, 2009.

\end{thebibliography}
\bibliographystyle{unsrt}

\end{document}